\documentclass[twocolumn, 	
 showpacs, 			
 preprintnumbers, 		
 aps,  				
 prd,  				
 a4paper, 			
 superscriptaddress, 		
 nofootinbib, 			
 tightenlines, 			
 floats 			
 ]{revtex4}

\usepackage{graphicx}
\usepackage{amssymb,latexsym}
\usepackage[draft=false]{hyperref}

\begin{document}

\renewcommand{\topfraction}{0.99}

\title{Cosmological parameter estimation and the inflationary
cosmology} 

\author{Samuel M.~Leach}
\email{s.m.leach@sussex.ac.uk} 
\affiliation{Astronomy Centre, University of Sussex, 
 Brighton BN1 9QJ, United 
Kingdom}

\author{Andrew R.~Liddle}
\email{a.liddle@sussex.ac.uk}
\affiliation{Astronomy Centre, University of Sussex, 
 Brighton BN1 9QJ, United 
Kingdom}

\author{J\'er\^ome Martin}
\email{jmartin@iap.fr}
\affiliation{Institut d'Astrophysique de Paris, 
 98bis boulevard Arago, 75014 Paris, France}

\author{Dominik J.~Schwarz}
\email{dschwarz@hep.itp.tuwien.ac.at}
\affiliation{Institut f\"ur Theoretische Physik, 
 Technische Universit\"at Wien,
 Wiedner Hauptstra\ss e 8--10, 1040 Wien, Austria} 
\date{\today} 

\pacs{98.80.Cq, 98.70.Vc \hfill astro-ph/0202094}


\begin{abstract}
We consider approaches to cosmological parameter estimation in the inflationary 
cosmology, focussing on the required accuracy of the initial power spectra. 
Parametrizing the spectra, for example by power-laws, is well suited to testing 
the inflationary paradigm but will only correctly estimate cosmological 
parameters if the parametrization is sufficiently accurate, and we investigate 
conditions under which this is achieved both for present data and for upcoming 
satellite data. If inflation is favoured, reliable estimation of its physical 
parameters requires an alternative approach adopting its detailed predictions. 
For slow-roll inflation, we investigate the accuracy 
of the predicted spectra at first and second order in the slow-roll expansion
(presenting the complete second-order corrections for the tensors for 
the first time). We find that 
within the presently-allowed parameter space, there are regions where it will be 
necessary to include second-order corrections 
to reach the accuracy requirements of \emph{MAP} and \emph{Planck} satellite 
data. 
We end by proposing a data analysis pipeline appropriate for testing inflation 
and for cosmological parameter estimation from high-precision data.
\end{abstract}

\maketitle

\section{Introduction}

Recent cosmic microwave background (CMB) anisotropy results
\cite{newCMB}, showing a multiple peak structure in the anisotropy
power spectrum, lend powerful support to the inflationary cosmology as
the origin of structure in the Universe. It is now widely expected
that cumulative improvements in the CMB data will lead to
progressively more accurate estimation of cosmological parameters,
with projects funded so far culminating in the \emph{Planck} satellite
mission expected to report results around 2010.
 
Given a set of data on structures in the Universe, such as the CMB
power spectrum, it is necessary to simultaneously fit both for the
parameters describing the global cosmology (such as the matter budget
and expansion rate) and those describing the so-called `initial
perturbations'; they cannot be considered separately. If the model
for the initial perturbations is insufficiently accurate, or even
worse completely wrong, the full power of the experiment to constrain
cosmological parameters cannot be exploited.
 
The inflationary cosmology is an attractive paradigm for the
generation of the initial perturbations, but even there the
situation can be very complicated in general. In particular, if there are
multiple scalar fields the perturbations can be a mixture of
isocurvature and adiabatic, and may be non-gaussian. Such initial
conditions may prove difficult or even impossible to parametrize, and
if such an inflation model is correct it will be a major obstacle to
successful parameter estimation. However it remains a powerful
working hypothesis that the simplest class of models, where inflation
is driven by a single scalar field, is viable; this creates a 
framework within which the necessary calculations are reasonably
simple, with the initial perturbations computed either approximately
analytically or exactly numerically. As yet, there is no indication
from observations that we might need to go beyond this paradigm.
 
The main goal of this article is to investigate different strategies
that an observer can use to estimate the cosmological parameters, and
to examine the extent to which it is necessary to adopt detailed
inflationary predictions. The spectrum of the fluctuations   is
assumed to be produced by an underlying inflationary model and is
calculated exactly by means of numerical computations. Given this
situation, we study how the data analysis can be performed in two
different scenarios. The first scenario applies if one wants only
to estimate cosmological parameters, such as the baryon density and
reionization optical depth, and does not care about the underlying
inflation model beyond being confident that the description of the
initial perturbations used is adequate. In this case, observers
typically use a power-law fit, see {\it e.g.}~Ref.~\cite{newCMB}, and
the first question is to test how accurate a power-law fit is to
typical inflationary cosmologies. In particular, we wish to know if
this kind of fit is accurate enough for present data, and whether it
will also be accurate enough to analyze high-precision data like that
to be provided by the \emph{Planck} satellite. The second scenario,
which makes more stringent requirements on theoretical accuracy, is if
one intends to estimate properties of the inflationary model. In this
case, the slow-roll method can be used to calculate an approximate
spectrum and we will study its accuracy. We also consider to which
order in the slow-roll parameters the spectrum should be calculated in
order to reach the \emph{Planck} precision. We propose an analysis
pipeline for testing the consistency of single-field slow-roll
inflation and estimating physical parameters of inflation, {\it e.g.}
the energy scale of inflation.
 
\section{Inflationary basics}
\label{sec:basics}

A single-field inflation model generates Gaussian spectra of purely
adiabatic density perturbations (scalar perturbations) and
gravitational waves (tensor perturbations). We denote the
dimensionless power spectra by ${\cal P}_{\cal R}(k)$, where ${\cal
R}$ is the intrinsic curvature perturbation on comoving hypersurfaces
(identical with Bardeen's $\zeta$ \cite{Bardeen89} up to a sign), and
${\cal P}_h(k)$, $h$ being the amplitude of gravitational
waves. Scalar and tensor perturbations obey the equation of a
parametric oscillator \cite{Muk}
\begin{equation}
\label{eqmotion}
\mu _{\rm S,T}''+
\biggl[k^2-\frac{z_{\rm S,T}''}{z_{\rm S,T}}\biggr]\mu _{\rm S,T}=0,
\end{equation}
where a prime denotes differentiation with respect to conformal time
$\eta$ and $k$ is the comoving wavenumber. This equation only requires
the assumption of linear perturbation theory. The quantities $\mu
_{\rm S,T}(\eta )$ are defined by $\mu _{\rm S}(\eta )\equiv 2z_{\rm
S}{\cal R}$ and $\mu _{\rm T}(\eta )\equiv z_{\rm T}h$ where \mbox{$z_{\rm
S}\equiv a\sqrt{2-aa''/a'^2}$} and $z_{\rm T}=a$. The initial conditions
for the mode functions $\mu _{\rm S,T}$ are fixed by the assumption
that the quantum fields are in the vacuum state when the mode $k$ is
subhorizon,
\begin{equation}
\label{ini}
\lim_{k/aH\rightarrow \infty }\mu _{\rm S,T}(\eta )=
\frac{4\sqrt{\pi }}{m_{\rm Pl}}\;
\frac{e^{-ik(\eta -\eta _{\rm i})}}{\sqrt{2k}},
\end{equation}
where $\eta _{\rm i}$ is an arbitrary initial 
time at the beginning of inflation. The power spectra are calculated 
according to 
\begin{equation}
\label{defspec}
{\cal P}_{{\cal R}}(k)= \frac{k^3}{8\pi ^2}
\biggl\vert\frac{\mu_{\rm S}}{z_{\rm S}}
\biggr \vert ^2, \quad {\cal P}_h(k)= \frac{2k^3}{\pi ^2}\biggl \vert 
\frac{\mu _{\rm T}}{z_{\rm T}}\biggr \vert ^2.
\end{equation}
Both power spectra can be derived from the inflaton potential
$V(\phi)$ and the initial conditions for the inflaton field $\phi $,
and hence are not independent. They can be obtained numerically by
solving the appropriate mode equations wavenumber by wavenumber (see
{\it e.g.}~Ref.~\cite{GL1}). In the following, they are denoted by
${\cal P}_{\rm num}(k)$.
 
The tensor-to-scalar ratio 
\begin{eqnarray}
\label{eqn:R}
R &\equiv& {{\cal P}_h\over {\cal P}_{\cal R}},
\end{eqnarray}
is of interest for testing the consistency of a given model of
inflation. It has often been defined in terms of the microwave
background quadrupole moments. This definition has the disadvantage
that it depends on the cosmological parameters, especially the density
of the cosmological constant $\Omega_\Lambda$ \cite{Knox,MRS}. In
Ref.~\cite{WTZ} the ratio between ${\cal P}_\Phi$ (where $\Phi$ is the
gauge-invariant Bardeen metric potential \cite{Bardeen80}) and ${\cal
P}_h$ was used, which removes the dependence on
$\Omega_\Lambda$. However $\Phi$ does still depend on the dynamics of
the Universe at the photon decoupling epoch and thus is not completely
model independent (it depends mainly on the physical matter density
$\omega_{{\rm m}} \equiv \Omega_{{\rm m}}h^2$).  The advantage of
${\cal R}$ is that it is conserved on super-horizon scales once the
decaying mode is negligible and provided only adiabatic perturbations
are considered \cite{Bardeen89,MS,Wands}.
 
The spectral indices and their ``running'' are defined by the following 
expressions
\begin{eqnarray}
\label{defn}
n_{\rm S}(k) -1 &\equiv & \frac{{\rm d}\ln {\cal P}_{{\cal R}}}{{\rm d} \ln k},
\quad 
n_{\rm T}(k) \equiv \frac{{\rm d} \ln {\cal P}_h}{{\rm d} \ln k},
\\ 
\alpha _{\rm S}(k) &\equiv & \frac{{\rm d}n_{\rm S}}{{\rm d}\ln k}, \quad 
\alpha _{\rm T}(k) \equiv \frac{{\rm d}n_{\rm T}}{{\rm d}\ln k}.
\end{eqnarray}
 
For purposes of illustration, in this paper we use three qualitatively
different inflationary models to mimic idealized measurements of the
power spectra. The first is a chaotic inflation model with a quartic
potential \cite{Linde83}
\begin{equation}
V(\phi)=\lambda \phi^4,
\end{equation}
the second a false vacuum inflation potential
\begin{equation}
V(\phi)=
V_0\biggl[1+\frac{1}{2}\mu^2\left(\frac{\phi}{m_{\rm Pl}}\right)^2\biggr],
\end{equation}
with $\mu^2=1$, which is inspired by the scenario of hybrid inflation 
\cite{hybrid}, and the third a potential introduced in 
Ref.~\cite{WMS}
\begin{equation}
V(\phi) = V_0\biggl[1-\frac{2}{\pi }\mbox{arctan }
\biggl(5\frac{\phi }{m_{\rm Pl}}\biggr)\biggr] \,.
\end{equation}
For each potential we need to know the scalar field value $\phi_*$
when observable perturbations were generated ({\it i.e.}~when a given
scale $k_*$ was equal to the Hubble radius during inflation),
corresponding roughly to $55$ $e$-foldings from the end of
inflation. The last two potentials provide no natural end to
inflation, and we make an arbitrary choice for $\phi_*$ to be equal  
$0.3\sqrt{2} \,m_{\rm Pl}$ and $-0.3m_{\rm Pl}$ respectively. The chaotic
inflation model ends by violation of slow-roll and so we take $\phi_*
\simeq 4.2m_{\rm Pl}$. 

\begin{figure}[t]
\includegraphics[width=\linewidth]{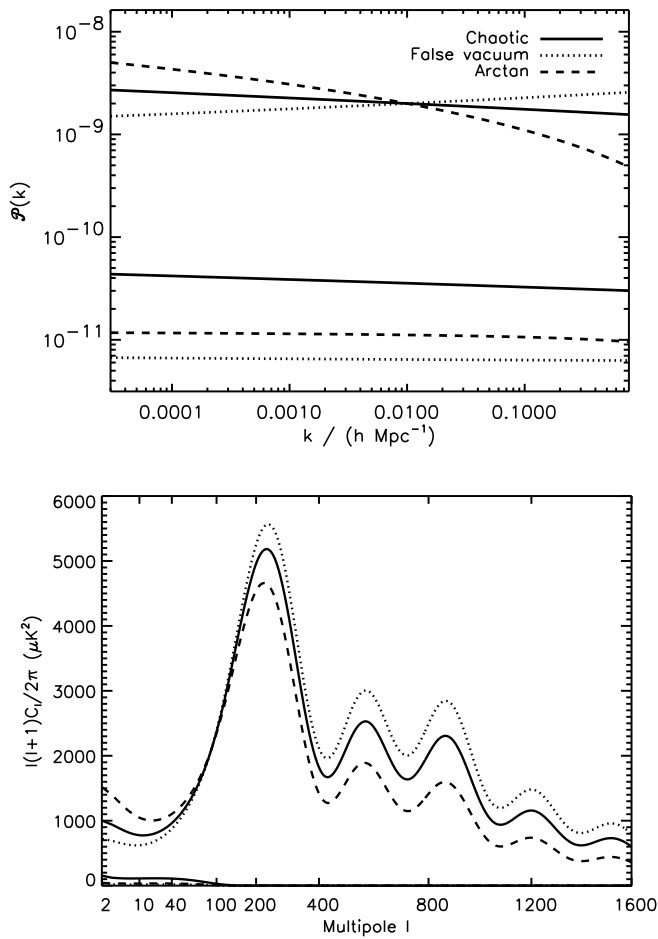}\\
\caption[fig0]{\label{fig:plot_num} The top panel shows the power
spectra of scalar (upper lines) and tensor (lower lines) perturbations
for our three models. The scalar spectra are normalized to ${\cal
P}_{{\cal R}} = 2\times10^{-9}$ at the scale $k_* =
0.01h\mbox{Mpc}^{-1}$, which approximately matches the COBE
normalization.  The bottom panel shows the corresponding $C_{\ell}$
curves for a flat cosmological model with $\omega_{{\rm b}}=0.0200$,
$\omega_{{\rm m}}=0.1268$ and $\omega_{\Lambda}=0.2958$ (implying
$h=0.65$), and reionization optical depth $\tau = 0.05$, with the
upper lines again the scalar contribution and the tensors considerably
subdominant. Only the sum of the two can be detected, though they
contribute differently to polarization anisotropies.}
\end{figure}

Fig.~\ref{fig:plot_num} shows the scalar and tensor power spectra
${\cal P}_{\rm num}(k)$ for these three models, obtained
numerically by the method of Ref.~\cite{GL1}. The corresponding microwave 
anisotropies, obtained using
a modified version of {\sc camb} \cite{LCL}, are also shown.\footnote{A module 
to directly input the predictions of slow-roll inflation to the {\sc camb} 
program is available to download at {\tt 
www.astronomy.susx.ac.uk/$\sim$sleach/inflation/}} We
present the characteristic quantities of these spectra, evaluated at
$k_*$, in Table~\ref{table1}. We have chosen these three models
because the chaotic model is an example in which tensor perturbations
are relevant and it shows moderate negative tilt, the false vacuum
model has a moderate positive tilt and the arctan model has both large
tilt and running.

\begin{table}
\begin{tabular}{l c c c c c}
Exact values & $R$ & $n_{\rm S} -1$ & $n_{\rm T}$ & $\alpha_{\rm S}$ 
& $ \alpha_{\rm T}$ \\ 
\hline \hline 
Chaotic & 0.285 & -0.055 & -0.037 & -0.0009 & -0.0006 
\\
False vacuum & 0.051 & \ 0.054 & -0.006 & \ 0.0018 & \ 0.0005
\\
Arctan model & 0.089 & -0.216 & -0.015 & -0.0298 & -0.0036
\\
\hline \hline
\end{tabular}
\caption{\label{table1}Numerical values of spectral indices, their 
running and the tensor-to-scalar ratio for the three models considered. All 
quantities are evaluated at $k_* =0.01h\mbox{Mpc}^{-1}$.}
\end{table}

\section{Parameter estimation ignoring inflationary predictions}
\label{sec:ignoreinf}

\subsection{Parametrizing the spectra}

To estimate the cosmological parameters we need an adequate
parametrization of scalar perturbations, and more sophisticated
analyses informed by inflation also include tensors.\footnote{A
general analysis would also have to consider vector modes and the
various possible isocurvature modes, but at present there is no
evidence that they are required.} If one is only interested in a
measurement of those cosmological parameters that do not describe the
initial perturbations, one would like to know whether robust results
can be obtained using simple forms for the initial power spectra
rather than detailed inflationary predictions. Therefore, in the context
envisaged in this section, the observer does not use the assumption
that inflation is the correct underlying theory, other than to
motivate the restriction of the scalar perturbations to be adiabatic.
 
It is common practice to assume a power-law shape for the spectrum
specified by an amplitude and a spectral index. The reasoning for this
parameterization is its simplicity. In the absence of any physical
model for the generation of fluctuations, one assumes that there is no
distinguished physical scale in the primordial power spectra. In order
to allow for mildly scale-dependent power spectra, a running of the
spectral indices can be included. This leads to the following shape
\begin{equation}
\label{explogex}
{{\cal P}_{\rm fit}(k)={\cal P}_{\rm fit}(k_*)}\left(k\over k_*\right)^{
n_{\rm fit}+\frac{1}{2}\alpha _{\rm fit}\ln (k/k_{*})},
\end{equation}
where $n_{\rm fit }$ is either $n_{\rm S}-1$ or $n_{\rm
T}$.\footnote{Note from the definition of the spectral index that
$n_{\rm S}(k)-1 \neq n_{\rm fit}+\frac{1}{2}\alpha _{\rm fit}\ln
(k/k_{*})$ away from the pivot scale.}  The pivot scale $k_*$ is the
scale at which all the quantities are evaluated.   A useful way of
viewing Eq.~(\ref{explogex}) is that it is the first terms of a Taylor
expansion of $\ln {\cal P}(k)$ in $\ln k$ about the pivot scale, which
draws one's attention to the possibility of using other expansions.

The simplest assumption would be to take both spectra as constant
(scale-invariant). Models with $n_{\rm S} - 1 = 0$ and $n_{\rm T} = 0$
do in fact provide acceptable fits to recent CMB data (for
sufficiently low or zero tensor amplitude); thus, if we decide to
ignore inflation for the moment, there is no reason from CMB
observations alone to include a tilt. Ref.~\cite{WTZ} quotes $n_{{\rm
S}} - 1 = -0.07^{+0.75}_{-0.16}$ at $95\%$ confidence level, and the
addition of large-scale structure data greatly tightens the constraint
without altering the conclusion that $n_{{\rm S}} = 1$ is
allowed. Current observational constraints on the running of the
spectral index are far weaker than the magnitude predicted in popular
inflationary models. Thus far, only upper limits on the contribution
of gravitational waves have been derived
\cite{upperR,WTZ,2dFCMB}. Some of these limits suffer from the
problems described below Eq.~(\ref{eqn:R}), and use strong priors on
some of the other cosmological parameters. Translating the result of
Fig.~5 of Ref.~\cite{WTZ} ($r<0.5$ at $95\%$ confidence level) to our
notation gives $R = 9r/25 < 0.2$, while Ref.~\cite{2dFCMB} gives a
weaker constraint also consistent with $R=0$. Let us also remark that
the majority of recent papers estimating parameters from the microwave
background have done so under the assumption that the scalar spectrum
has a power-law shape and that there is no contribution from tensor
perturbations ($R=0$)

The question of how far power spectra expansions should be taken, and
how accurately their coefficients need to be computed, obviously
depends on the accuracy and dynamic range of observations. For present
observations an accuracy level of ten percent or better is certainly
required.  Ultimately, \emph{Planck} will measure multipole moments
$C_\ell$ from $\ell$ of 2 to about 2000, corresponding to $\Delta \ln
k/k_* \simeq 3.5$ on either side of a central pivot $k_*$. It is
rather unclear how accurately the multipole moments need to be
represented at the extremes (cosmic variance intervening on large
scales and the damping tail removing the signal on short scales), but
in the centre an accuracy of better than one percent is certainly
desired (see {\it e.g.}~Ref.~\cite{TE}).\footnote{We note that current 
implementations of {\sc cmbfast} \cite{SZ} and {\sc camb} \cite{LCL} have a 
target accuracy of one percent, so there is presently nothing to gain by 
demanding power spectrum accuracy much higher than this.} If one then further 
assumes
that \emph{Planck} data will be combined with high-accuracy galaxy
correlation data, the $k$-range might extend to around $\Delta \ln
k/k_* \simeq 6$ (corresponding to $k_{{\rm max}} \simeq 30 h\ {\rm
Mpc}^{-1}$), though the nonlinearly-evolved galaxy power spectrum on
short scales is unlikely to be amenable to extremely accurate
multi-parameter estimation. The choice of pivot scale $k_*$ is
important as the difference between the fitted and the true power
spectrum produces an error that runs as we move away from the pivot
scale. While a careful tracking of error covariances should lead to
results independent of the choice of pivot, those covariances should
be minimized to a good approximation by aligning $k_*$ with $\ell_*$,
the multipole where we expect the observational errors to be
least. One can use the approximation \cite{Huthesis}
\begin{equation}
k_*=\frac{H_0}{2} \frac{\sqrt{\Omega_{\rm m}}}{1+0.084
\ln\Omega_{\rm m}}\ell_*,\;\;\;\;\Omega_{\rm m}+\Omega_{\Lambda}=1,
\end{equation}
where $H_0=h/3000\,{\rm Mpc}^{-1}$ to carry out this alignment. 
 
Having described what are the typical errors in the multipole moments,
Error($C_{\ell }$), we now need to link this quantity to the error in
the power spectrum itself, Error(${\cal P}$), since this is the
quantity calculated in practice. We assume throughout this paper that
an error in our determination of the power spectrum propagates
directly to an error in our determination of the $C_\ell$'s since
\begin{equation}
C_\ell = 4\pi \int {\rm d}\ln k \;{\mathcal P}(k)[\Delta_\ell(k)]^2,
\end{equation}
where $\Delta_\ell(k)$ is the $\ell$-th momentum of the temperature
fluctuations. In other words, we assume that Error($C_{\ell
}$)$\simeq$ Error(${\cal P}$).

Another question is how an error in the power spectrum propagates to
an error in the estimation of the cosmological parameters. In
general the Fisher matrix formulation is needed to estimate how well a
given experiment can measure the parameters; the error in
the cosmological parameters is not simply related to the error in
the power spectrum as there are many
parameters and lots of degeneracies amongst them.  The requirement
Error(${\cal P}$)$\simeq 1\%$ for \emph{Planck} is a very stringent
condition. In particular, it does not imply that parameter
estimates would go astray if we drifted outside our power spectrum
accuracy criterion; we would expect parameter estimates to stabilize some
way before the fitted power spectrum was within our $1\%$ accuracy
everywhere. Our criterion is a sufficient and conservative condition
for establishing a safe procedure: as long as the power spectrum
accuracy is below $1\%$ everywhere, we are confident that the systematic
errors coming from an inaccurate parametrization of the initial
conditions will not play a role in the data analysis of an experiment
like \emph{Planck}. 

\subsection{Accuracy of the parametrized spectra}
 
We now investigate the systematic errors which might arise from
assuming that the spectra have perfect power-law shapes or, in a more
sophisticated version, a constant value for the running of the
spectral index.  The first step is to fix the numerical values of the
coefficients ${\cal P}_{\rm fit}(k_*)$, $n_{\rm fit}$ and $\alpha
_{\rm fit}$. We have no means to calculate them theoretically in the
present context.  In practice, observers determine these coefficients
by carrying out a fit to the data. Here, we carry out a least-squares
fit of ${\cal P}_{\rm fit}(k)$ to ${\cal P}_{\rm num}(k)$ to obtain
best-fit scale-invariant, power-law and power-law plus running
spectra. This means that the coefficients ${\cal P}_{\rm fit}(k_*)$,
$n_{\rm fit}$ and $\alpha _{\rm fit}$ are those for which the quantity
\begin{equation}
\sum_i \biggl[{\cal P}_{\rm fit}(k_i)-{\cal P}_{\rm num}(k_i)\biggr]^2
\end{equation}
is minimized. We took the $k_i$ to be equally spaced in ${\rm d} \ln
k$ and given equal weight.  This idealized fitting approach will tend
to sacrifice accuracy in the centre of desired range in favour of
accuracy at the extremes. Here the idea is to test whether in
principle the shape of $P_{\rm fit}$ can reproduce the true power
spectrum over a reasonable range in $k$. This obviously becomes
important if, for example, we try to use a power-law shape to fit to a
model with significant running of the spectral index.  The result of
the minimization procedure for the three models introduced above is
summarized in Table~\ref{table2}. These values should be compared with
the exact ones of Table~\ref{table1}.

\begin{table}
\begin{tabular}{l c c c c c c}
Fitted values & $A_{\rm fit/num}$
& $R$ & $n_{\rm S} -1$ & $n_{\rm T}$ & $\alpha_{\rm S}$ & 
$\alpha_{\rm T}$ \\
\hline \hline 
Chaotic & 1.05 & 0.279 & & & & 
\\
 & 1.00 & 0.285 & -0.054 & -0.036 & & 
\\
 & 1.00 & 0.285 & -0.055 & -0.037 & -0.0010 & -0.0007
\\
False vacuum & 0.98 & 0.053 & & & & 
\\
 & 1.01 & 0.051 & \ 0.054 & -0.006 & & 
\\
 & 1.00 & 0.051 & \ 0.054 & -0.006 & \ 0.0017 & \ 0.0004
\\
Arctan model & 1.23 & 0.072 & & & & 
\\
 & 0.95 & 0.092 & -0.178 & -0.016 & & 
\\
 & 0.99 & 0.090 & -0.238 & -0.020 & -0.0303 & -0.0044
\\
\hline \hline
\end{tabular}
\caption{\label{table2}The ratio of the fitted amplitude to the
numerical amplitude of the scalars, $A_{\rm fit/num}$, and the
best-fit values of the tensor-to-scalar ratio, the spectral indices
and their running for the three models considered at
$k_*=0.01h\mbox{Mpc}^{-1}$. For each model, we present the results for
a scale-invariant, power-law and power-law with running spectral shape
in three rows respectively.}
\end{table}

For the first two examples we conclude that the sequence of fitting a
constant amplitude, a power-law, and finally a power-law with running
provides best-fit values which reproduce the numerical values with
sufficient accuracy. From the observational point of view this is
reflected in the fact that the best-fit value of $R$ is the same in
the second and third row and that the fit values of the spectral
indices are almost the same as well.  Such a behavior is the
experimental evidence that the input does make sense.  The situation
is different for our third example, the arctan model.  Although there
is slow convergence in the best-fit values of $R$, no sign of
convergence can be detected by inspection of Table~\ref{table2} in the
spectral indices. This is confirmed by a comparison with the numerical
values of Table~\ref{table1}, {\emph e.g.}, the fitted spectral index
of the tensors is less precise in the third row than in the second,
the scalar spectral index is underestimated by $0.036$ by the
power-law fit and overestimated by $0.022$ including running. From the
point of view of inflationary parameters, see below, these are large
fluctuations. Thus an observer in possession of sufficiently-accurate
data sometime in the future should conclude for our third example that
more parameters have to be introduced in the fit, before any physical
meaning can be extracted from the best-fit values of the spectral
indices. 

A large error in the fitted values of the amplitude and of the
spectral index, even if we are not interested in the physics of
inflation for the moment, is undesirable for two reasons. Firstly the
overall amplitude of scalar perturbations is a quantity that we hope
to measure from \emph{MAP} and \emph{Planck} at the percent level. Thus we
would prefer to relate it to a physical quantity, namely the amplitude
of superhorizon density fluctuations. The second reason comes from
considerations of large-scale structure data, where it is customary to
include a linear bias parameter, $b$, to account for the overall
normalization of the matter power spectrum. If we simultaneously fit
to the CMB, then we can only assign any physical meaning to $b$ if we
are certain that the amplitude of scalar perturbations is correct. In
addition, an inaccurate estimate of the amplitude and the tilt could
spoil a consistency check of structure formation based on measurements
of $\sigma_8$.

Having determined the coefficients, the second step is now to compute
the error. We define this by
\begin{equation}
\mbox{Error}({\cal P}) \equiv \left({{\cal P}_{\rm fit}\over
   {\cal P}_{\rm num}} -1\right)\times 100 \% \,. 
\end{equation}
In the following, we give three examples.

\begin{figure}[t]
\includegraphics[width=\linewidth]{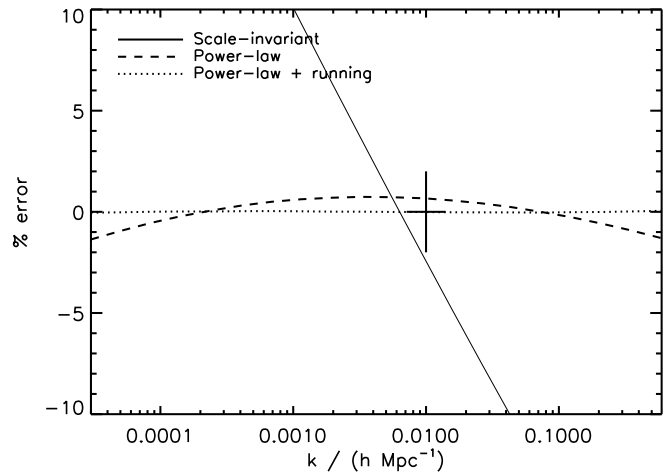}\\
\caption[fig1]{\label{fig:errorS_fit_fv} Error curves for
various fits to the scalar power
spectrum for the false vacuum model. While the power-law fit is acceptable 
for fitting to present data, neglecting running affects the estimate of
the power spectrum amplitude at the pivot point at the percent
level.}
\end{figure}

In Fig.~\ref{fig:errorS_fit_fv} we plot the error in the scalar power
spectrum in the case of the false vacuum model. The best-fit
scale-invariant spectrum is a poor fit for this particular model,
while the best-fit power-law spectrum improves things greatly, keeping
the errors below $2\%$ which is more than adequate for present CMB
data and marginally adequate for \emph{Planck}. The large effect of
the tilt is due to a long lever arm in wavenumbers \cite{MS2}; the
error being of the order $(n_{\rm S}-1)\ln k/k_*$, even a small tilt
can have a significant effect if the data span several decades in
wavenumbers. We can further see that with the inclusion of running,
$P_{\rm fit}$ now reproduces the power spectrum in great detail. This
is because the correction to the spectrum is of order $\alpha_{\rm S}
\ln^2 k/k_*$ which, for the running of this example of $\alpha_{\rm
S}\simeq 0.002$, gives a significant effect though the correction is
much smaller than that from the tilt.

Given a set of observations, the importance of running is tested by
including it in the fit and examining whether the fit improves
significantly. In the absence of any theoretical prejudice, one might
well hope to detect significant running at high significance. However,
some of the simplest inflation models predict running of at least an
order of magnitude below what even \emph{Planck} can achieve
\cite{CGL}. In that case there will be no significant detection of
running, and marginalizing over the running permitted by the
observations may lead to a significant inflating of errors on other
parameters.  While combining short-scale observations with the
microwave background may give a stronger lever-arm in constraining
running, this may well turn out to be a parameter for which it is
desirable to investigate imposing a strict theoretically-motivated
prior to compare with a free fit.  Further, even if running is
detected at high significance this problem then resurfaces concerning
the running-of-running.

\begin{figure}[t]
\includegraphics[width=\linewidth]{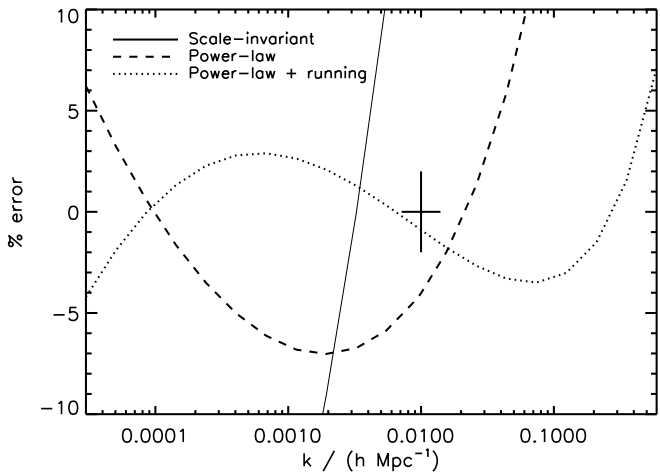}\\
\caption[fig2]{\label{fig:errorS_fit_wang} As 
Fig.~\ref{fig:errorS_fit_fv} but for the arctan model. For 
fitting to data of the present quality, the inclusion of the running is 
required.} 
\end{figure}

As we have already concluded from the discussion of the best-fit
values in Table~\ref{table2}, there exist models for which a power-law
fit to the spectrum does not provide a good description in contrast to
the above example. In Fig.~\ref{fig:errorS_fit_wang} the error in the
scalar power spectrum for the arctan model is displayed. Including the
running is necessary for the present accuracy of CMB experiments.  Now
the effect of running is comparable to that of the tilt. We see that
more parameters ({\it e.g.}~running of the running) would be necessary
to reproduce the power spectrum with $1\%$ accuracy. We actually have
to add more and more parameters until the spectrum starts to converge (see also 
the
discussion of Table~\ref{table2}), or consider using a different spectral shape.

\begin{figure}[t]
\includegraphics[width=\linewidth]{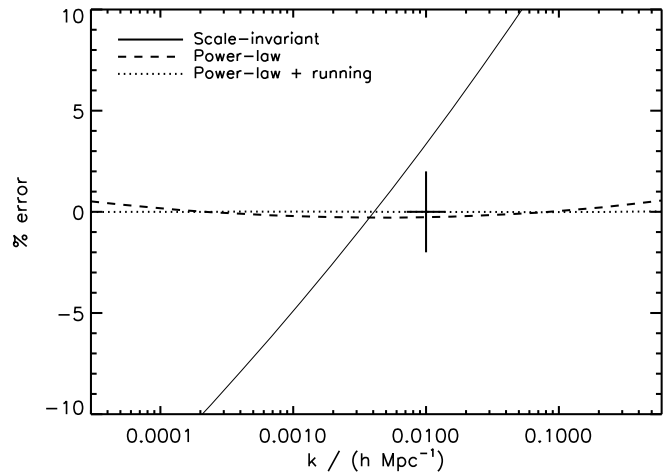}\\
\caption[fig3]{\label{fig:errorT_fit_lambda55} As 
Fig.~\ref{fig:errorS_fit_fv} but for the chaotic inflation model
tensor spectrum. Although the percentage error is large for the
scale-invariant fit, the absolute error is small compared to the
scalar spectrum, and so the scale-invariant fit is still acceptable.}
\end{figure}

For the tensors in the case of the chaotic model, we see in
Fig.~\ref{fig:errorT_fit_lambda55} that the spectrum is poorly fitted
by the scale-invariant spectrum. However, the accuracy requirements on
the tensor spectrum are less stringent --- the tensor amplitude is
generally less that the scalar amplitude and so the required absolute
error in ${\cal P}_{{\rm h}}$ is also less [$\simeq
R\times\mbox{Error}({\cal P}_{\cal R})$]. Thus, a typical inflationary
tensor spectrum is well described even by a scale-invariant spectrum
for present-day experiments, though there is no reason not to describe
it with the same sophistication as the scalar spectrum. For future CMB
measurements, the inclusion of a tilt is sufficient in this example.
 
To answer the main question of this section, we can expect to obtain
robust estimates for the cosmological parameters for a restricted
class of inflationary models using the fitting procedure described
above. However, there exist models where this is no longer true. In
the following sections, we specify the criteria which define this
class of models.

\section{Predictions of slow-roll inflation}
\label{sec:srinf}

In this section, we restrict our considerations to the class of
slow-roll models of inflation. The advantage is that we can now
predict the shape of the power spectra and link the parameters
characterizing these spectra to the physics of inflation. There has
recently been renewed progress in the accurate calculation of
inflationary perturbations by analytical techniques, including a
computation of the power spectra to arbitrary order in the slow-roll
expansion for single-field inflation by Stewart and Gong \cite{SG},
and a computation at higher order for models that may violate one of
the slow-roll conditions \cite{STG,S}. We utilize the Stewart--Gong
results here as they have the most general applicability, extending
them with an explicit evaluation of higher-order terms for the tensor
spectrum.
 
The background evolution can be described in terms of the horizon-flow
parameters $\{\epsilon_n\}$ \cite{STG}. Starting from $\epsilon_0
\equiv H(N_{\rm i})/H(N)$, where $1/H$ is the Hubble distance and $N
\equiv \ln (a/a_{\rm i})$ the number of $e$-folds since some initial
time $t_{\rm i}$, the set $\{\epsilon_n\}$ is defined by
\begin{equation}
\label{flow}
\epsilon_{n+1} \equiv {{\rm d} \ln |\epsilon_n|
\over {\rm d} N}, \qquad n\geq 0.
\end{equation} 
These parameters can be easily related to various definitions of the
slow-roll parameters. Setting $n = 1$ we find $\epsilon_1 = -{\rm
d}\ln H/{\rm d}\ln a$, which is nothing but the slow-roll parameter
$\epsilon$ of Refs.~\cite{LPB,RMP}.  The parameter $\eta$ of
Refs.~\cite{LPB,RMP}, which is usually defined to measure the
deceleration of the inflaton field, enters as $\epsilon_2 = 2\epsilon
- 2\eta$. The third slow-roll parameter, $\xi$, is contained in
$\epsilon_2 \epsilon_3 = 4 \epsilon^2 - 6 \epsilon \eta + 2 \xi^2$. In
this notation, all the $\epsilon_n$ are typically of the same order of
magnitude.  Inflation takes place provided $\epsilon_1 < 1$. Slow-roll
inflation is defined by the condition $|\epsilon_n| \ll 1$, for all
$n>0$.
 
A measurement of the horizon-flow parameters, at a specific moment
during inflation, would immediately provide us with a value for the
inflaton potential $V$ and its derivatives with respect to the
inflaton field $\phi$ (denoted by a prime in what follows) for any
single-field inflation model. For example, from $H$ and
$\epsilon_1,\epsilon_2$, and $\epsilon_3$ we can calculate the
potential and its first two derivatives exactly,
\begin{eqnarray}
V &=& \frac{3 m_{\rm Pl}^2 H^2}{8 \pi} \left(1 - \frac{\epsilon_1}{3}\right),\\
V' &=& -\frac{3 m_{\rm Pl} H^2}{(4 \pi)^{1/2}} \, \epsilon_1^{1/2}
 \left(1 - \frac{\epsilon_1}{3} + \frac{\epsilon_2}{6}\right), \\
\frac{V''}{3H^2} &=& 2 \epsilon_1 - \frac{\epsilon_2}{2}
	-\frac{2\epsilon_1^2}{3} +
 \frac{5\epsilon_1 \epsilon_2}{6} -\frac{\epsilon_2^2}{12} - 
\frac{\epsilon_2 \epsilon_3}{6} \,.
\end{eqnarray}
If $\epsilon_3$ cannot be determined and the horizon-flow parameters
are small compared to unity, we can still estimate $V''$ by keeping
the leading terms only.
 
For slow-roll models we can invert this procedure and estimate the
horizon-flow parameters. At leading order in these parameters we find:
\begin{eqnarray}
H^2 &\simeq& \frac{8\pi}{3 m_{\rm Pl}^2} V, \\
\epsilon_1 &\simeq& \frac{m_{\rm Pl}^2}{16 \pi} \left({V'\over V}\right)^2,\\
\epsilon_2 &\simeq& \frac{m_{\rm Pl}^2}{4 \pi} 
 \left[\left({V'\over V}\right)^2 - {V''\over V}\right], \\
\epsilon_2 \epsilon_3 &\simeq& \frac{m_{\rm Pl}^4}{32 \pi^2} \left[
 {V''' V'\over V^2} - 3{V''\over V}\left({V'\over V}\right)^2\! 
 + 2 \left({V'\over V}\right)^4\right]\! . \ \ \ \ 
\end{eqnarray} 
To give an example, for chaotic inflation with the potential $V
\propto \phi^\gamma $ we find $\epsilon_1 \simeq \gamma /4 \Delta N$
and $\epsilon_2 \simeq \epsilon_3 \simeq 1/\Delta N$, where $\Delta N$
denotes the number of $e$-folds before inflation ends. Chaotic
inflation is a simple model where the higher horizon-flow parameters
are of the same order of magnitude as lower ones. In the case of
power-law inflation ($a \propto t^p$) where the potential is given by
$V \propto \exp[-(16\pi/p)^{1/2}\phi/m_{\rm Pl}]$, we recover the
exact result $\epsilon_1 = 1/p$ and $\epsilon_2 = \epsilon_3 = 0$.

The power spectra of scalar and tensor perturbations can be obtained
approximately using analytic techniques. One expands the power spectra
about some particular wavenumber $k_*$, and then computes the
coefficients using the slow-roll expansion or some other scheme of
approximation. This amounts to a double approximation. Given that we
need to cover several orders of magnitude in $k$, the most appropriate
expansion variable is $\ln k$, giving
\begin{equation} 
\label{logex}
{{\cal P}(k)\over{\cal P}_0(k_*)} = a_0 + a_1 \ln \left(k\over k_*\right) 
 + \frac{a_2}{2} \ln^2\left(k\over k_*\right)
 + \dots 
\end{equation}
The next step is to establish an expression for the coefficients
$a_n$, which can be obtained either with help of the slow-roll
expansion \cite{SL,LPB,LL,MS2,MRS,SG} or the methods of approximation
developed in Refs.~\cite{STG,S}. Since the former covers a more
general class of inflation models then the latter, we focus on
slow-roll inflation in the following. We will use the term {\em
first-order} to refer to results including all terms up to order
$\epsilon_m$ and {\em second-order} if one goes to terms including
$\epsilon_m^2$.
 
The normalization of the power spectra is set by the expansion rate
during inflation, $H$, and the parameter $\epsilon_1$, namely
\begin{eqnarray}
\label{eqn:pnorm}
{\cal P}_{{\cal R}0}(k_*) &=& {H^2 \over \pi \epsilon_1 m_{\rm Pl}^2}, \\ 
{\cal P}_{h0}(k_*) &=& {16 H^2\over \pi m_{\rm Pl}^2},
\end{eqnarray} 
where $H$ and $\epsilon_1$ are evaluated when $aH = k_*$ during
inflation.  The scalar amplitude has been calculated up to first-order
in the slow-roll parameters by Stewart and Lyth \cite{SL}, and
recently up to second-order by Stewart and Gong \cite{SG}. These
calculations are sufficient to allow calculation of an infinite,
though incomplete, set of expansion coefficients of which the first
few are given by
\begin{eqnarray}
\label{eqn:as0}
a_{{\rm S}0} &=& 1 - 2\left(C + 1\right)\epsilon_1 - C \epsilon_2 
 + \left(2C^2 + 2C + {\textstyle\frac{\pi^2}{2}} - 5\right) 
 \epsilon_1^2 \nonumber \\
 & & + \left(C^2 - C + {\textstyle\frac{7\pi^2}{12}} - 7\right) 
 \epsilon_1\epsilon_2 
 + \left({\textstyle\frac 12} C^2 + {\textstyle\frac{\pi^2}{8}} 
 - 1\right)\epsilon_2^2 \nonumber \\
 & & + \left(-{\textstyle\frac 12} C^2 
 + {\textstyle\frac{\pi^2}{24}}\right)\epsilon_2\epsilon_3 , \\
a_{{\rm S}1} &=&
 - 2\epsilon_1 - \epsilon_2 + 2(2C+1)\epsilon_1^2 
 + (2C - 1)\epsilon_1\epsilon_2 \nonumber \\
 & & + C\epsilon_2^2 - C\epsilon_2\epsilon_3 ,\\
a_{{\rm S}2} &=& 
 4\epsilon_1^2 + 2\epsilon_1\epsilon_2 + \epsilon_2^2 - 
\epsilon_2\epsilon_3 ,
\label{eqn:as2}
\end{eqnarray}
where $C \equiv \gamma_{\rm E} + \ln 2 - 2 \approx -0.7296$. For the
tensors, the corresponding set is as follows:
\begin{eqnarray}
a_{{\rm T}0} &=&
 1 - 2\left(C + 1\right)\epsilon_1 
 + \left(2C^2 + 2C + {\textstyle\frac{\pi^2}{2}} - 5\right) 
 \epsilon_1^2 \nonumber \\
 & & + \left(-C^2 - 2C + {\textstyle\frac{\pi^2}{12}} - 2\right) 
 \epsilon_1\epsilon_2 ,\\
a_{{\rm T}1} &=& 
 - 2\epsilon_1 + 2(2C + 1)\epsilon_1^2 
 - 2(C + 1)\epsilon_1\epsilon_2 ,
\label{eqn:at1} \\
a_{{\rm T}2} &=& 4\epsilon_1^2 - 2\epsilon_1\epsilon_2 .
\label{eqn:at2}
\end{eqnarray}
We have presented for the first time the ${\cal O}(\epsilon_n^2)$
terms in the tensor amplitude which we obtained along the lines of
Ref.~\cite{SG}.

The coefficients $a_n$ for $n>0$ can also be obtained by successive
differentiation of the first term of the expansion
\begin{eqnarray}
a_n &\equiv& 
 \left.{{\rm d}^n [{\cal P}(k)/{\cal P}_0(k_*)]\over {\rm d}\ln^n k}
 \right\vert_{k = k_*} \\
\label{diff}
 &=& \frac{1}{{\cal P}_0(k_*)}\left({1\over 1 -\epsilon_1}
	{{\rm d}\over {\rm d}N}\right)^n {\cal P}_0(k_*)a_0(k_*),
\end{eqnarray}
where we used the ``horizon crossing'' condition $k_* = k = aH$ to
obtain the second line. From Eqs.~(\ref{flow}) and (\ref{diff}) we see
that the leading contribution to $a_n$ is of order $\epsilon _m^n$
(where $\epsilon _m^n$ means any terms containing $n$ of the $\epsilon
$, not necessarily all the same). If $a_0$ has been written to first
order, differentiation yields $a_1$ to second order, $a_2$ to third
order and so on. Note that the coefficients of the Taylor series,
Eq.~(\ref{logex}), always feature an increasing number of powers of
the slow-roll parameters, so in practice convergence of the Taylor
series is governed by the size of $\epsilon_m \ln k/k_*$, which in
principle needs to be small for all values of $m \geq 1$. Thus the
series can still be strongly convergent even if $\ln k/k_*$ exceeds
one, as it will for typical upcoming experiments.
 
\begin{table}
\begin{tabular}{l c c c c c}
Slow-roll values & $R$ & $n_{\rm S} -1$ & $n_{\rm T}$ & $\alpha _{\rm S}$ 
& $ \alpha _{\rm T}$ \\ 
\hline \hline 
Chaotic & 0.285 & -0.055 & -0.037 & -0.0010 & -0.0007
\\
False vacuum & 0.051 & \ 0.054 & -0.006 & \ 0.0017 & \ 0.0004
\\
Arctan model & 0.089 & -0.221& -0.014 & -0.0291 & -0.0041
\\
\hline \hline
\end{tabular}
\caption{\label{table3}Slow-roll values of spectral indices, their 
running and the tensor-to-scalar ratio for the three models considered. All 
quantities are evaluated at $k=0.01h\mbox{Mpc}^{-1}$.}
\end{table}

Let us now calculate the spectral indices and their running in the 
slow-roll approximation up to second order. For this purpose, it 
useful to calculate the logarithm of the power spectrum
\begin{equation} 
\label{plex}
\ln {{\cal P}(k)\over{\cal P}_0(k_*)} = b_0 + b_1 \ln \left(k\over k_*\right)
+ \frac{b_2}{2} \ln^2\left(k\over k_*\right) + \dots .
\end{equation}
Exponentiation of Eq.~(\ref{plex}) automatically enforces the positive
definiteness of ${\cal P}(k)$ and allows us to directly link the first
coefficients $b_n$ to the spectral indices and the runnings, because
\begin{equation}
b_{{\rm S}1}=n_{\rm S} -1,\ b_{{\rm T}1}=n_{\rm T}, \ 
b_{{\rm S}2}=\alpha _{\rm S}, \ b_{{\rm T}2}=\alpha _{\rm T}. 
\end{equation}
The equivalent expressions to Eqs.~(\ref{eqn:as0}) -- (\ref{eqn:at2})
are
\begin{eqnarray}
\label{eqn:bs0}
b_{{\rm S}0} &=& 
 - 2\left(C + 1\right)\epsilon_1 - C \epsilon_2 
 + \left(- 2C + {\textstyle\frac{\pi^2}{2}} - 7\right) 
 \epsilon_1^2 \nonumber\\
 & & + \left(- C^2 - 3C + {\textstyle\frac{7\pi^2}{12}} - 7\right) 
 \epsilon_1\epsilon_2 
 + \left({\textstyle\frac{\pi^2}{8}} - 1\right) 
 \epsilon_2^2 \nonumber \\
 & & + \left(-{\textstyle\frac 12}C^2 + {\textstyle\frac{\pi^2}{24}}\right) 
 \epsilon_2\epsilon_3 ,\\
b_{{\rm S}1} &=& - 2 \epsilon_1 - \epsilon_2 - 2 \epsilon_1^2 
 - (2C + 3) \epsilon_1 \epsilon_2 
 - C \epsilon_2 \epsilon_3 ,\\
b_{{\rm S}2} &=& - 2 \epsilon_1 \epsilon_2 - \epsilon_2 
\epsilon_3\label{eqn:bs2}, 
\end{eqnarray}
for the scalars, and 
\begin{eqnarray}
b_{{\rm T}0} &=& 
 - 2\left(C + 1\right)\epsilon_1 
 + \left(- 2C + {\textstyle\frac{\pi^2}{2}} - 7\right) 
 \epsilon_1^2 \nonumber\\
 & & + \left(-C^2 - 2C + {\textstyle\frac{\pi^2}{12}} - 2\right) 
 \epsilon_1\epsilon_2 , \\
b_{{\rm T}1} &=& - 2\epsilon_1 - 2\epsilon_1^2 
 - 2(C + 1)\epsilon_1\epsilon_2 , \\
b_{{\rm T}2} &=& - 2\epsilon_1\epsilon_2,
\label{eqn:bt2}
\end{eqnarray}
for the tensors.
 
Finally, the ratio of amplitudes of scalars and tensors at the pivot
point is
\begin{eqnarray}
R &=& 16\epsilon_1 \left[1 
+ C \epsilon_2 + 
 \left(C- {\textstyle\frac{\pi^2}{2}} +5\right)\epsilon_1\epsilon_2 
 \right. \nonumber \\
 & & + \left.
 \left({\textstyle\frac12}C^2 - {\textstyle\frac{\pi^2}{8}} +1\right)
 \epsilon_2^2 
 + \left({\textstyle\frac12}C^2 - {\textstyle\frac{\pi^2}{24}}\right) 
 \epsilon_2 \epsilon_3 \right]. \ \ \ 
\label{eqn:Rsr}
\end{eqnarray}
This becomes the well-known ``consistency condition of inflation'' $R
= - 8 n_{\rm T}$ at leading order, which holds for
single-inflaton-field slow-roll models.  The values of the ratio $R$,
the spectral indices and their running, computed in the slow-roll
approximation for the three models envisaged in this article, are
summarized in Table~\ref{table3}. The values of the horizon-flow
parameters were obtained numerically, though an actual reconstruction
may also feature a slow-roll approximation in relating those to the
inflationary potential.

\section{Does the shape of the fitted spectra matter?}
\label{sec:shape}

In the preceding section, we have shown that the shape of the slow-roll spectra
does not coincide with the shape of the fit of Section III.  From a theoretical
point of view, it is clear that the former should be used not only to predict
the spectra but also to fit real data.  For many choices of parameters the
difference between the shapes is not significant, but there are also models
where this difference can be important.

\begin{figure}[t]
\includegraphics[width=\linewidth]{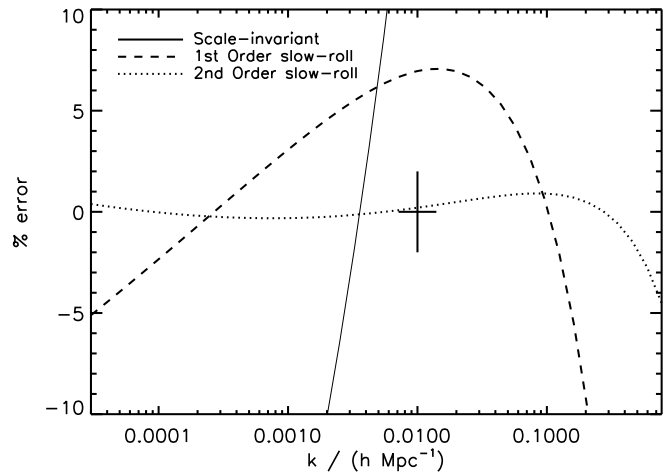}\\
\caption[fig5]{\label{fig:errorS_fitsr_wang} Fitting the slow-roll 
shape to the arctan model. The errors should be compared with the errors
in Fig.~\ref{fig:errorS_fit_wang}. For this model, the second-order
slow-roll shape provides a better fit that the power-law plus running
shape.}
\end{figure}

An example is given in Fig.~\ref{fig:errorS_fitsr_wang}, where we plot 
\begin{equation}
\mbox{Error}({\cal P}^{\rm fit}_{\rm sr}) \equiv 
\left({{\cal P}_{\rm sr}^{\rm fit}\over {\cal P}_{\rm num}} - 1\right)\times 
100 \%,
\end{equation}
for the arctan model of Section II. In this equation, 
${\cal P}^{\rm fit}_{\rm sr}$ is found by considering 
\begin{equation}
\label{eqn:psrc}
{\cal P}_{\rm sr} = c_0+c_1\ln \biggl(\frac{k}{k_*}\biggr )
+\frac{c_2}{2}\ln ^2\biggl(\frac{k}{k_*}\biggr ),
\end{equation} 
and calculating the three coefficients $c_0$, $c_1$ and $c_2$ by 
minimizing the quantity 
\begin{equation}
\sum_i \biggl[{\cal P}^{\rm fit}_{\rm sr}(k_i)-
{\cal P}_{\rm num}(k_i)\biggr]^2.
\end{equation}
Comparing the slow-roll fit of Fig.~\ref{fig:errorS_fitsr_wang} with
the power-law fit of Fig.~\ref{fig:errorS_fit_wang}, we can see that
the slow-roll shape does indeed provide a better fit in this case,
keeping the error below $1\%$ for most of the range. Thus the power
spectrum shape can make a difference, and there exist models where
fitting with the power-law instead of the slow-roll shape can lead to
significant errors (  defined by the criterion of Sec.~III-A  ).
 
However, one cannot conclude that the slow-roll shape necessarily
gives a better fit in general. An example where the slow-roll fit
converges slower than the power-law fit is the chaotic model, although
the difference is not significant in that case. For power-law
inflation the slow-roll shape will actually fare less well.

\begin{table}
\begin{tabular}{l c c c c c c}
Slow-roll fit & $A_{\rm fit/num}$
& $R$ & $n_{\rm S} -1$ & $n_{\rm T}$ & $\alpha _{\rm S}$ & 
$ \alpha _{\rm T}$ \\ 
\hline \hline 
Chaotic 	& 1.05 & 0.279 & & & & 
\\
 & 1.01 & 0.283 & -0.056 & -0.037 & & 
\\
 & 1.00 & 0.285 & -0.055 & -0.037 & -0.0010 & -0.007
\\
False vacuum & 0.98 & 0.053 & & & & 
\\
 & 1.01 & 0.051 & \ 0.051 & -0.006 & & 
\\
 & 1.00 & 0.051 & \ 0.055 & -0.006 & \ 0.0017 & \ 0.0004
\\
Arctan model & 1.23 & 0.072 & & & & 
\\
 & 1.07 & 0.082 & -0.210 & -0.016 & & 
\\
 & 1.00 & 0.089 & -0.213 & -0.019 & -0.0289 & -0.0044
\\
\hline \hline
\end{tabular}
\caption{\label{table4} As in Table~\ref{table2}, but for the 
spectral shape that is predicted by slow-roll inflation.}
\end{table}

A second step is to go from the coefficients $c_0$, $c_1$ and $c_2$ to
the characteristic parameters of the primordial spectra. This can be
done by means of the relations
\begin{eqnarray}
(n_{\rm S}-1)^{\rm fit}_{\rm sr} &=& \frac{c_1}{c_0}+{\cal O}(\epsilon_n^3), 
\\
(\alpha _{\rm S})^{\rm fit}_{\rm sr} &=& \frac{c_2}{c_0}-\frac{c_1^2}{c_0^2}
+{\cal O}(\epsilon_n^3),
\end{eqnarray}
and analogous equations for the tensors. The coefficient $R$ can be
obtained as
\begin{equation}
R^{\rm fit}_{\rm sr} =\frac{c_{\rm 0 T}}{c_{\rm 0S}}.
\end{equation}
The results are summarized in Table~\ref{table4}. This table should be
compared with Tables~\ref{table1} and \ref{table2}.  Fitting a
different shape has now the effect that the parameters of the arctan
model converge, in contrast to the power-law fit.

Fitting the coefficients $c_n$ allows us to test the consistency
relation of inflation, and thereafter constraining $c_{\rm 1T}$ and
$c_{\rm 2T}$ according to Eqs.~(\ref{eqn:at1}) and (\ref{eqn:at2})
allows us to measure the inflationary parameters.
 
Having shown that there exist situations where the shape matters, we
wish to find the region of the parameter space in which the difference
between a power-law shape with running and the shape predicted by
slow-roll inflation is significant. For this purpose, we define the
estimator
\begin{eqnarray}
\sigma &\equiv & \frac{{\cal P}_{{\rm sr}}-{\cal P}_{{\rm fit}}}{\left({\cal 
P}_{{\rm sr}}+{\cal P}_{{\rm fit}}\right)/2} \; \times 100\% \,; \\
 &\simeq& -\frac{n}{2}\biggl(\alpha +\frac{n^2}{3}\biggr)
\ln ^3\biggl(\frac{k}{k_*}\biggr)\times 100\% \,,
\label{eqn:sigmaapprox}
\end{eqnarray}
where $n$ stands in for $n_{{\rm S}}-1$ or $n_{{\rm T}}$. Note that
this estimator presumes that the two fits generate the same values for
the amplitude, spectral index and running, whereas in practice a
different choice of shape will lead to different values.  This
estimator therefore underestimates the differences between the two
fits close to the pivot point and overestimates them far away from the
pivot.

\begin{figure}[t]
\includegraphics[width=\linewidth]{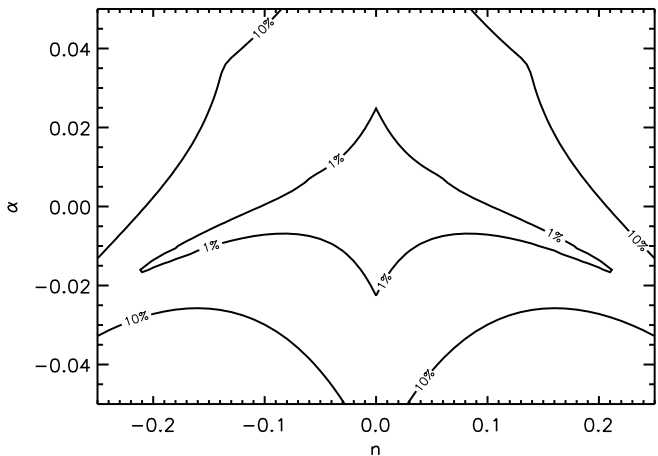}\\
\caption[fig6]{\label{fig:srshape} The region of fitted spectral indices and
runnings in which the difference between the power-law shape and the slow-roll
shape, estimated by $|\sigma|$, is within $1\%$ and within $10\%$.}
\end{figure}

In Fig.~\ref{fig:srshape} we plot the contours of the maximum of
$|\sigma(k)|$ in the interval $-1.5<\log_{10}(k/k_*)<1.5$ in the
$(n_{\rm S} -1,\alpha_{\rm S})$ plane. Its shape can be understood
most easily from the approximation Eq.~(\ref{eqn:sigmaapprox}).  We
conclude that within the ranges $n_{\rm S} -1 \in [-0.05,0.05]$ and
$\alpha_{\rm S} \in [-0.015,0.015]$, shape should not matter even at
the accuracy level of \emph{Planck}. For present CMB experiments this
plot suggests that as long as $|n_{\rm S} -1|$ is within the range
shown in Fig.~\ref{fig:srshape} shape is not an issue if the running
is at most of order $0.01$, which is the case for a wide class of
inflationary models (similar constraints should be assumed to hold
true for higher corrections as well).

A significant difference between the two fits at a given observational
accuracy is a clear indicator that higher-order terms may be
important, as it is those which give the difference between the two
expansions. To be certain of robust results, an attempt should be made
to estimate these higher-order terms, either by extending one or both
expansions to seek convergence between them or by resorting to fully
numerical analysis techniques.

\section{Accuracy of slow-roll analytic spectra}
\label{sec:accuracy}

In the previous section we showed that spectral shape can matter and
therefore that it is important to take the predictions of slow-roll
inflation into account if we are interested in the physics of
inflation itself. Before discussing how to extract the inflationary
parameters we study the accuracy of the slow-roll approximation at
second order. First studies of the accuracy of the slow-roll expansion
can be found for the amplitudes in Ref.~\cite{GL1} by comparing to
numerical results, while in Ref.~\cite{MS2} the first-order
expressions for the amplitudes and the spectral indices has been
tested by comparison to analytical results for power-law
inflation. Here we extend these studies to the full power spectrum at
second order.  We define the error of the slow-roll power spectrum as
\begin{equation}
\mbox{Error}({\cal P}) \equiv \left({{\cal P}_{\rm sr}\over
   {\cal P}_{\rm num}} - 1\right)\times 100\%, 
\end{equation}
where ${\cal P_{\rm sr}}$ is given by Eqs.~(\ref{logex}) and 
(\ref{eqn:as0})--(\ref{eqn:at2}). In these expressions the values of $H$, 
$\epsilon_1$, $\epsilon_2$ and $\epsilon_3$ are computed numerically for 
the three models of Sec.~II.

\begin{figure}[t]
\includegraphics[width=\linewidth]{plot_errorST_lambda55.epsi}\\
\caption[fig1]{\label{fig:errorST_lambda55} Scalar and tensor error
curves for the chaotic inflation potential.  The pivot scale crosses
the Hubble horizon $55$ $e$-folds before the end of inflation. We see
an improvement in accuracy from the first to the second-order
expressions. The tensors have better overall accuracy than the
scalars.}
\end{figure}

Looking at the chaotic inflation model first, we can see from
Fig.~\ref{fig:errorST_lambda55} that the error curves resulting from
slow-roll predictions generally have the property that they are most
accurate close to the pivot point (in terms of amplitude and spectral
index) and that the error increases as we move away from the pivot
point.  We can also see that the second-order expressions can improve
the accuracy of both the scalar and tensor power spectra to within
\emph{Planck} requirements, whereas the accuracy of the corresponding
first-order expression would be at best marginal. This improvement is
mostly brought about by the inclusion of the running.

The tensor spectrum of Fig.~\ref{fig:errorST_lambda55} is determined
more accurately than the scalars. We have observed that this is
typically the case.  Since the accuracy requirement upon the tensors
is less than on the scalars, it is the scalars upon which attention
should be focused.

\begin{figure}[t]
\includegraphics[width=\linewidth]{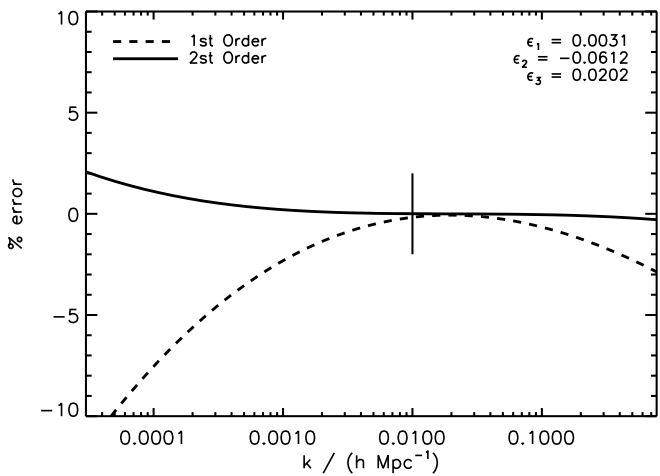}\\
\caption[fig1]{\label{fig:errorS_fv}
Scalar error curve for the false vacuum inflation model. Again, we see
an improvement in accuracy from the first to second-order expressions
which helps to correct the amplitude at the pivot point.}
\end{figure}

Next we turn to the false vacuum inflation model. Note immediately
from Fig.~\ref{fig:errorS_fv} that the second-order expression
improves both the shape of the power spectrum and the accuracy of the
amplitude at the pivot point itself. The first-order expression is
good enough for present experiments in this example, but not for \emph{MAP}
and \emph{Planck}. 

\begin{figure}[t]
\includegraphics[width=\linewidth]{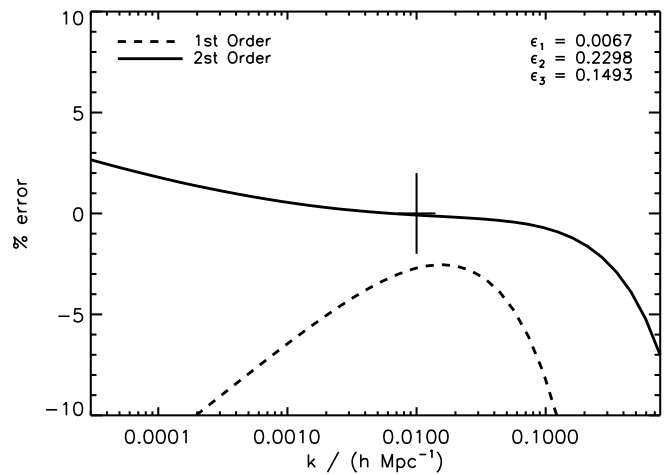}\\
\caption[fig1]{\label{fig:errorS_wang}
Scalar error curve for the arctan potential. We see an improvement
from the first to second order as well as a significant correction to
the overall amplitude at the pivot scale.}
\end{figure}

Finally, for the arctan model we see in Fig.~\ref{fig:errorS_wang}
that although $\epsilon_1$ is small and $\epsilon_2$ and $\epsilon_3$
are still in agreement with the slow-roll conditions, the effect of
the second-order correction is very important. The first-order
expression is not sufficient for \emph{MAP}. In this example, the first-order
expression also produces a significant error in the amplitude at the
pivot point.  For \emph{Planck} the plot suggests that the third order
is necessary.

It is of course impossible to study the accuracy of all possible
models of inflation in this way. We therefore need a more general
estimator for the accuracy of the slow-roll expansion in the parameter
space $\epsilon_n$.  The difference between the slow-roll expansions
of ${\cal P}(k)$ and $\ln {\cal P}(k)$ is such an estimator. We define
the error at a given order $n$ to be
\begin{equation}
\label{sign}
\sigma_{{\rm n}}= 
\frac{\left\vert\sum_{i=0}^n {a_i\over i!} \ln^i\!\!\left(\frac{k}{k_*}\right)-
\exp\left[\sum_{i=0}^n{b_i\over i!}
\ln^i\!\!\left(\frac{k}{k_*}\right)\right]\right\vert}{\sum_{i=0}^n 
{a_i\over i!}\ln^i\!\!\left(\frac{k}{k_*}\right) +
\exp\left[\sum_{i=0}^n{b_i\over i!}
\ln^i\!\!\left(\frac{k}{k_*}\right)\right]} \times 100\% \,, 
\end{equation}
where the coefficients $a_i$ and $b_i$ are taken at order
$\epsilon^n_m$. The interpretation of this expression is that it gives
the smallest fractional amount by which the worse of the two
expansions departs from the true power spectrum, namely half the
distance between the two estimates.   This interpretation justifies
the absence of a factor $1/2$ at the denominator in
Eq.~(\ref{sign}).  

This expression is of order $\epsilon_m^{n+1}$ and therefore is an
indicator of the importance of orders that have not been
included. Moreover it has the same typical behaviour of the errors as
one goes away from the pivot point, and we also find that it estimates
the orders of the errors for the examples of Sec.~II correctly. We
expect that this estimate typically works well although there
exists the possibility of fine-tuning models such that
$\sigma_n$ is not a good estimator. In the following we study the
maximum of the error in a suitable interval of wavenumbers, because a
large error in a small range may spoil an otherwise
accurate fit. We therefore maximize
$\sigma_1(k,\epsilon_1,\epsilon_2)$ and
$\sigma_2(k,\epsilon_1,\epsilon_2,\epsilon_3)$ over
$-1.5<\log_{10}(k/k_*)<1.5$. This is certainly conservative but is a
good indicator of when robust results are expected.

\begin{figure}[t]
\includegraphics[width=\linewidth]{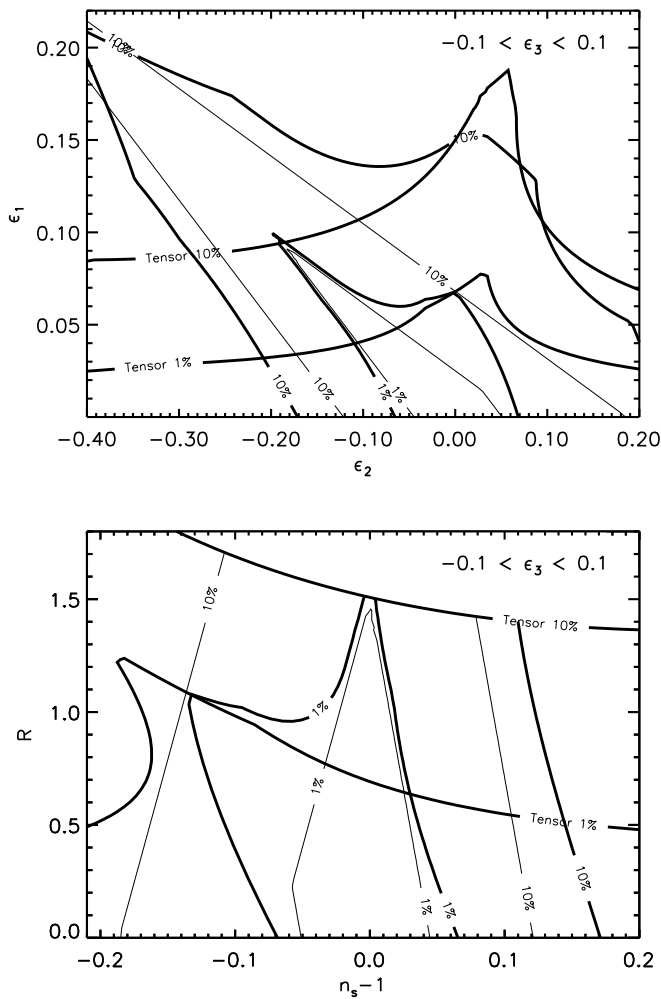}\\
\caption[fig1]{\label{fig:e1e2} These panels show the error estimate
$\sigma_1$ for the slow-roll expansions at first order (thin lines)
and $\sigma_2$ at second order (thick lines). The upper panel is as a
function of horizon-flow parameters, while the lower panel transforms
this into the $(n_{{\rm S}}-1)$--$R$ plane.}
\end{figure}

The upper panel of Fig.~\ref{fig:e1e2} shows the error in the
$\epsilon_1$--$\epsilon_2$ plane, maximizing over
$-0.1<\epsilon_3<0.1$ (the arctan model actually lies outside this
range).  The scalar error contours are elongated along the direction
$\epsilon_1=-\epsilon_2/2$, which corresponds to $n_{\rm S} = 1$ at
first order.  In the top left corner $\sigma_1$ becomes independent of
the dominant contribution proportional to $\ln k$ for $n_{\rm S} = 1$.
For $\sigma_2$ there is a similar cancellation of the $\ln^2 k$
contribution for models close to $n_{\rm S} = 1$, which explains the
shape of the contours.  These elongated shapes are therefore a feature
of our estimator $\sigma_n$; they do not reflect a proper estimate of
the error in the top left corner as other higher-order terms not
considered would spoil these cancellations.  With the exception of
that top region, we see that, as expected, the second-order
expressions extend the area of parameter space meeting a specified
accuracy requirement.

It is useful to examine these results in the $(n_{{\rm S}}-1)$--$R$
plane via the transformation
\begin{eqnarray}
n_{\rm S} -1 &=& - 2 \epsilon_1 - \epsilon_2, \\
R &=& 16\epsilon_1 \label{1stcons}
\end{eqnarray}
shown in the lower panel of Fig.~\ref{fig:e1e2}. We use the
first-order relations also for the second-order error contours here;
the error made by this can be neglected for the present purpose. The
restriction that we put on $\epsilon_3$ gives rise to values for
running in the range $\alpha_{\rm S} \in [-0.023,0.14]$ for the
displayed region of parameter space. The first-order expression gives
errors within 10\% in the region given approximately by $-0.15<n_{{\rm
S}}-1<0.1$ and $R<1.5$. The second-order slow-roll expression gives an
accuracy better than 1\% in a somewhat smaller range of parameter
space ($-0.1<n_{{\rm S}}-1<0.05$, $R<1.0$).

It is important to stress that these regions are very conservative as
we maximize the error over both $\epsilon_3$ and wavenumber. The
conclusions of small errors in parameter space regions is therefore
very robust, and indeed the errors are likely to be within acceptable
levels even for many models lying outside our contours.

\begin{figure}[t]
\includegraphics[width=\linewidth]{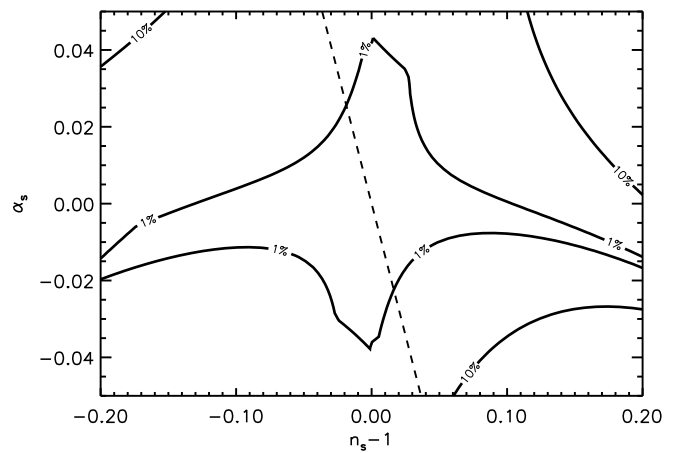}\\
\caption[fig1]{\label{fig:nsdnsdlnk} The error estimate $\sigma_2$ in
the $(n_{\rm S}-1)$ -- $\alpha_{\rm S}$ plane, with $\epsilon_1 \ll
0.001$. The dashed line is $\alpha_{\rm S} = (n_{\rm S} - 1) / C$, in
the vicinity of which the error estimate can be misleading.}
\end{figure}

An important limit is when $\epsilon_1$ is very small, since a broad
class of inflation models belong to this category, {\emph e.g.}~false
vacuum dominated inflation gives rise to tiny $\epsilon_1$.  When
$\epsilon_1 \lesssim 0.001$, then the tensor spectrum will have no
effect on the low-$\ell$ portion of the $C_\ell$ curves at the $1\%$
level, see Eq.~(\ref{eqn:Rsr}). At this point the tensor $C_\ell$'s
drop out of reach and we can no longer measure $H$ during inflation
and $\epsilon_1$ separately, see Eq.~(\ref{eqn:pnorm}). The scalar
power spectrum, Eqs.~(\ref{eqn:as0}) -- (\ref{eqn:as2}), now reduces
to a function of ${\cal P}_{{\cal R}0}(k_*)$, $\epsilon_2$ and
$\epsilon_2\epsilon_3$, where the last two parameters determine
$n_{\rm S} - 1 = - \epsilon_2 - C \epsilon_2 \epsilon_3$ and
$\alpha_{\rm S} = - \epsilon_2 \epsilon_3$. In
Fig.~\ref{fig:nsdnsdlnk} we plot the error of the second-order power
spectrum, $\sigma_2$, in the $(n_{\rm S}-1)$ -- $\alpha_{{\rm S}}$
plane. The transformation between the $(n_{\rm S}-1)$ -- $\alpha_{{\rm
S}}$ plane and the $\epsilon_2$ -- $\epsilon_3$ plane is nonlinear and
singular at $\epsilon_2 = 0$ for any $\epsilon_3$. All corresponding
models have $n_{\rm S}-1 = \alpha_{\rm S} = 0$. Moreover, in the
vicinity of the line $\alpha_{\rm S} = (n_{\rm S} -1)/C$ the value of
$\epsilon_2$ becomes arbitrarily small, and thus $\epsilon_3$ can be
huge. Therefore, in the vicinity of the dashed line the estimator
$\sigma_2$ is misleading, because it gives a small error even for
models which violate the slow-roll condition $\epsilon_3 \ll
1$. Nevertheless, the conclusion is that fairly weak running $<0.02$
can be accurately ($1\%$) described by a slow-roll expansion with tiny
$\epsilon_1$.

\begin{figure}
\setlength{\unitlength}{\linewidth}
\begin{picture}(1,0.95)
\put(0.545,0.92){up to ${\mathcal O}_{{\rm pl}}$}
\put(0.5,0.9){\vector(0,-1){0.05}}
\put(0.72,0.8){\line(0,1){0.1}}
\put(0.72,0.9){\line(-1,0){0.22}}
\put(0.15,0.8){\makebox(0,0){\framebox{\parbox{1.8 cm}{CMB data}}}}
\put(0.27,0.8){\vector(1,0){0.105}}
\put(0.5,0.8){\makebox(0,0){\framebox{
  \parbox{1.8 cm}{convergence of pl+r fit?}}}}
\put(0.62,0.8){\vector(1,0){0.2}}
\put(0.85,0.8){\circle{0.05}}
\put(0.85,0.8){\makebox(0,0){$1$}}
\put(0.65,0.76){no}
\put(0.42,0.71){yes}
\put(0.545,0.71){up to ${\mathcal O}_{{\rm sr}}$}
\put(0.72,0.69){\vector(-1,0){0.22}}
\put(0.72,0.6){\line(0,1){0.09}}
\put(0.5,0.75){\vector(0,-1){0.10}}
\put(0.5,0.6){\makebox(0,0){\framebox{
  \parbox{1.8 cm}{convergence of sr fit?}}}}
\put(0.62,0.6){\vector(1,0){0.2}}
\put(0.85,0.6){\circle{0.05}}
\put(0.85,0.6){\makebox(0,0){$1$}}
\put(0.65,0.56){no}
\put(0.42,0.50){yes}
\put(0.5,0.55){\vector(0,-1){0.10}}
\put(0.5,0.4){\makebox(0,0){\framebox{\parbox{1.8 cm}{consistency of fits?}}}}
\put(0.62,0.4){\vector(1,0){0.2}}
\put(0.85,0.4){\circle{0.05}}
\put(0.85,0.4){\makebox(0,0){$1$}}
\put(0.65,0.36){no}
\put(0.42,0.29){yes}
\put(0.5,0.35){\vector(0,-1){0.125}}
\put(0.5,0.2){\makebox(0,0){\framebox{\parbox{1.8 cm}{$\Omega = 1$?}}}}
\put(0.62,0.2){\vector(1,0){0.2}}
\put(0.85,0.2){\circle{0.05}}
\put(0.85,0.2){\makebox(0,0){$1$}}
\put(0.5,0.175){\vector(0,-1){0.11}}
\put(0.65,0.16){no}
\put(0.42,0.11){yes}
\put(0.5,0.04){\circle{0.05}}
\put(0.5,0.04){\makebox(0,0){$2$}}
\end{picture}
\begin{picture}(1,0.6)
\put(0.15,0.55){\circle{0.05}}
\put(0.15,0.55){\makebox(0,0){$2$}}
\put(0.18,0.55){\vector(1,0){0.2}}
\put(0.5,0.55){\makebox(0,0){\framebox{\parbox{1.8 cm}{tensors?}}}}
\put(0.62,0.55){\vector(1,0){0.2}}
\put(0.85,0.55){\circle{0.05}}
\put(0.85,0.55){\makebox(0,0){$3$}}
\put(0.65,0.51){no}
\put(0.42,0.46){yes}
\put(0.5,0.52){\vector(0,-1){0.12}}
\put(0.5,0.35){\makebox(0,0){\framebox{\parbox{1.8 cm}{consistency check}}}}
\put(0.62,0.35){\vector(1,0){0.2}}
\put(0.85,0.35){\circle{0.05}}
\put(0.85,0.35){\makebox(0,0){$1$}}
\put(0.65,0.31){no}
\put(0.42,0.25){yes}
\put(0.5,0.3){\vector(0,-1){0.1}}
\put(0.5,0.15){\makebox(0,0){\framebox{\parbox{1.8 cm}{{\small \tt :-)} measure 
   $H,\epsilon_1,\epsilon_2, \epsilon_3$}}}}

\put(0.05,0.173){\circle{0.05}}
\put(0.05,0.173){\makebox(0,0){$1$}}
\put(0.1,0.163){physics beyond}
\put(0.1,0.123){slow-roll}
\put(0.05,0.075){\circle{0.05}}
\put(0.05,0.075){\makebox(0,0){$2$}}
\put(0.1,0.062){set $\Omega=1$, test sr inflation}
\put(0.05,0.01){\circle{0.05}}
\put(0.05,0.01){\makebox(0,0){$3$}}
\put(0.1,0.00){assume sr; upper bound on $\epsilon_1$ and $H$}

\end{picture}
\caption{Suggested pipeline to test slow-roll inflation and estimate 
its parameters.\label{fig:pipeline}}
\end{figure}
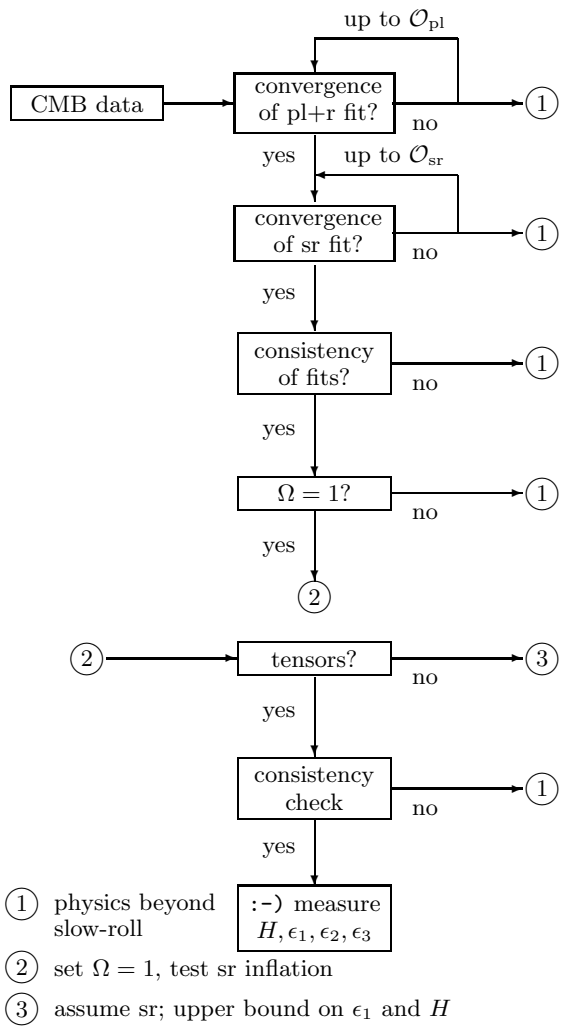

\section{Testing slow-roll inflation}

We end with a proposal of how to proceed with testing slow-roll
single-field inflation using future high-accuracy data. The corresponding 
data analysis pipeline is sketched in Fig.~\ref{fig:pipeline}. The inputs 
are the CMB data and a cosmological model (\emph{e.g.}~$\Lambda$CDM). The 
first step should be to determine the cosmological parameters under the 
assumption that the power spectra of scalar and tensor perturbations are 
given by a power-law with running of the spectral index, see 
Eq.~(\ref{explogex}).  One should check the convergence of the values
of all cosmological parameters as one fits scale-invariant, power-law,
and power-law with running spectra, as discussed in Sec.~\ref{sec:ignoreinf}.
One should continue to refine the power spectrum shape (adding in running
of running etc) until the new power spectrum parameter is found to be 
consistent with zero. At this point one has the choice to neglect this 
final parameter, and this seems a sensible option. We call the order of 
this truncated power spectrum ${\mathcal O}_{{\rm pl}}$. 
In a similar manner one should also check the convergence of the
cosmological parameter estimates while fitting to the data using 
scale-invariant, first-order and then second-order slow-roll shapes,
up to order ${\mathcal O}_{{\rm sr}}$.

One should find $|{\mathcal O}_{{\rm pl}}-{\mathcal O}_{{\rm sr}}|\leq
1$, with ${\mathcal O}_{{\rm pl}}={\mathcal O}_{{\rm sr}}$ being the
most likely case.  If we also find consistent estimates of the
cosmological parameters then clearly the choice of power spectrum
shape doesn't matter.  If ${\mathcal O}_{{\rm pl}}\neq {\mathcal
O}_{{\rm sr}}$ but the cosmological parameter estimates are convergent
and consistent with each other, then we have some evidence that a
particular power spectrum shape may be preferred.  Figure
\ref{fig:srshape} might be used to check whether the extracted
spectral indices and runnings are expected to give rise to a
significant difference between the two fits.

If there is no convergence using one or both of the power spectrum shapes,
or if the different power spectrum shapes lead to significantly different
estimates of the cosmological parameters, then there is either
a significant problem in the assumed cosmological model or the shape
of the spectrum is completely different from a power-law,
\emph{e.g.}~a pronounced bump or a step at a privileged scale
\cite{bump+step}.  Presuming the latter, within the context of
single-field inflation, the optimal strategy is a direct estimation of
the inflationary potential from the data itself, without using
intermediate approximations such as the slow-roll expansion, as
described by Grivell and Liddle \cite{GL2}.\footnote{The inflationary
potential is parametrized, for example by a Taylor series, and the
scalar and tensor power spectra are obtained by solving the mode
equations and fed into a Boltzmann code such as {\sc cmbfast}
\cite{SZ} or {\sc camb} \cite{LCL}.  The only approximation is the
validity of linear perturbation theory.  The result is an unbiased
estimation of the inflationary potential with automatic generation of
the error covariances of the potential parameters amongst themselves
and with the cosmological parameters \cite{GL2}. Other considerations of 
single-field inflation beyond slow-roll are given in Refs.~\cite{nonSR,STG}.}  
Such a 
calculation
must simultaneously fit all parameters, and so will also test whether
the results are consistent with a flat universe; the simplest models
of inflation predict $\Omega_{{\rm total}} = 1 \pm 10^{-5}$, though
realistic experiments will be orders of magnitude larger in
uncertainty.  If so the data are consistent with inflation, but
single-field slow-roll inflation would be ruled out.

If satisfactory convergence of the cosmological parameters is achieved
then the next step is to check whether $\Omega_{{\rm total}}$ is consistent 
with one. If this test is failed then slow-roll inflation is excluded and 
we need alternative physics. If the Universe is consistent with flatness, 
slow-roll inflation can now be taken very seriously. In the previous 
section we have shown that the power of fluctuations can be predicted
at the required level of accuracy in a large region of parameter space
favoured by present CMB observations. Once slow-roll inflation 
has been adopted as a working hypothesis, $\Omega_{{\rm total}}$ should be 
fixed at unity and not varied in any parameter fits.

We can now test the consistency relation and then estimate the inflationary 
parameters. In principle one could use either expansion [Eq.~(\ref{explogex}) 
or Eq.~(\ref{eqn:psrc})] if it has been successful, and even if 
${\mathcal O}_{{\rm pl}}\neq {\mathcal O}_{{\rm sr}}$ the inflationary
information contained within them should be equivalent. However
presuming it is available it makes best theoretical sense to use the
slow-roll fit. The approach is sketched in the lower tree of the
pipeline in Fig.~\ref{fig:pipeline}.

The first step is to check whether a tensor contribution can be
detected at a significant level. If not, then there are no means to
fully check the specific predictions of slow-roll inflation. However,
this means that an upper bound on the tensor-to-scalar ratio $R$ is
provided by the CMB data. Assuming slow-roll inflation we can use the
consistency relation Eq.~(\ref{eqn:Rsr}) [or its first-order version
Eq.~(\ref{1stcons})] to obtain an upper bound on $\epsilon_1$. Then we
neglect all $\epsilon_1$ terms in Eqs.~(\ref{eqn:as0}) --
(\ref{eqn:at2}), allowing an estimate of $\epsilon_2$, $\epsilon_3$
and the normalization of the scalar power spectrum $H^2/\pi \epsilon_1
m_{\rm Pl}^2$.  Together with the upper bound on $\epsilon_1$ this
gives an upper bound on the scale of inflation $H$. 
Figure \ref{fig:nsdnsdlnk} might be used to estimate the theoretical 
error in the measurement of $\epsilon_2$ and $\epsilon_3$. If the estimates
for $|\epsilon_2|$ and $|\epsilon_3|$ turn out to be larger than the
upper bound on $\epsilon_1$ we can take these estimates
seriously. However, if it turns out that one of the higher-order
parameters is of the same order as the upper bound for $\epsilon_1$ we
cannot consistently neglect $\epsilon_1$. In this case only a
banana-shaped region in parameter space of the second-order slow-roll
expansion can be identified. But a warning is required at that point;
without a detection of tensors it might be impossible to distinguish between 
single-field slow-roll inflation and other models.

If there is a significant detection of tensors, the next step is to
test the consistency equation of slow-roll inflation
Eq.~(\ref{eqn:Rsr}). If this test is not passed, we have ruled out
single-field slow-roll inflation. If we find consistency, the final
step is to measure the scale of inflation $H$ and the inflationary
parameters $\epsilon_1$, $\epsilon_2$ and $\epsilon_3$. By fitting
directly for these parameters, rather than the coefficients of
expansion as above, we are now automatically imposing the consistency
relations between the scalar and tensor spectra. This is also
important for measurement of the cosmological parameters, as it
ensures that the uncertainties are not overestimated (under the
presumption that slow-roll inflation is correct). The slow-roll shape
is the preferred option for carrying out this final parameter
determination, and this is also the determination which yields the
definitive measures of the various cosmological parameters. These
might differ from the parameters estimated from the power-law plus
running fit once the consistency conditions are imposed. In particular
the uncertainties should tighten as the inflationary predictions are
more specific than fitting free power-laws plus running. The systematic 
uncertainty from theory in the measurement of inflationary parameters
can now be estimated with the help of Fig.~\ref{fig:e1e2}. 

Having analytically reconstructed an inflationary potential, its
validity can be checked by evaluating the perturbations generated by
the potential numerically, which will provide a further estimate of
the magnitude of higher-order corrections. If these prove significant,
the numerical results could be used to `tune' the reconstructed
potential with the aim of removing any biases in estimation of other
parameters. Ultimately, analytic results obtained the way we describe
can be compared with a direct numerical reconstruction as described in
Ref.~\cite{GL2}, with the two methods providing invaluable
cross-checks on each other.

We have presented a strategy to measure the most important quantity 
in the context of inflationary models, the scale of inflation $H$. It  
probes the time scale and thus the energy scale of new physics, which 
requires the detection of tensor contributions. Sensitivity to gravitational 
waves is mainly provided via high-sensitivity polarization measurements, 
and it is these which may allow us to probe the highest energy scales for the
first time.

\section*{Acknowledgments}

We thank Lloyd Knox, Max Tegmark and C\'esar Terrero-Escalante for
useful discussions.  S.M.L.~was supported by PPARC and A.R.L.~in part
by the Leverhulme Trust.  D.J.S.~acknowledges a visit to the Sussex
Astronomy Centre funded by the Austrian Academy of Sciences, the Royal
Society and PPARC.

 

\end{document}